\documentclass[11pt,draftcls]{IEEEtran}

\usepackage{bbm,dsfont,mathrsfs,fixmath}
\usepackage{amsmath,amssymb,eucal,graphicx}
\usepackage{stmaryrd}
\usepackage{amsmath,amstext,amsfonts,amssymb}
\usepackage{epsfig}
\usepackage{exscale}

\newtheorem{definition}{Definition}
\newtheorem{lemma}{Lemma}

\newtheorem{prop}{Proposition}[section]

\newtheorem{assumption}{Assumption}[section]
\newtheorem{remark}{\indent \bf Remark}[section]
\newtheorem{property}{Property}[section]
\newtheorem{theorem}{Theorem}

\def\LSB{\left[}        
\def\RSB{\right]}       

\newcommand{\mb}{\mathbf}

\newcommand{\mc}{\mathcal}

\newfont{\bbb}{msbm10 scaled 500}

\newfont{\bb}{msbm10 scaled 1100}
\newcommand{\CC}{\mbox{\bb C}}

\newcommand{\av}{{\bf a}}

\newcommand{\gv}{{\bf g}}
\newcommand{\hv}{{\bf h}}

\newcommand{\nv}{{\bf n}}

\newcommand{\wv}{{\bf w}}
\newcommand{\vv}{{\bf v}}
\newcommand{\xv}{{\bf x}}
\newcommand{\yv}{{\bf y}}
\newcommand{\zv}{{\bf z}}

\newcommand{\onev}{{\bf 1}}

\newcommand{\Am}{{\bf A}}

\newcommand{\Gm}{{\bf G}}
\newcommand{\Hm}{{\bf H}}
\newcommand{\Id}{{\bf I}}

\newcommand{\Nm}{{\bf N}}

\newcommand{\Sm}{{\bf S}}

\newcommand{\Xm}{{\bf X}}
\newcommand{\Ym}{{\bf Y}}
\newcommand{\Zm}{{\bf Z}}

\newcommand{\Cc}{{\cal C}}

\newcommand{\Gc}{{\cal G}}
\newcommand{\Hc}{{\cal H}}

\newcommand{\Nc}{{\cal N}}

\newcommand{\Sc}{{\cal S}}
\newcommand{\Tc}{{\cal T}}

\newcommand{\Wc}{{\cal W}}

\newcommand{\Phim}{\hbox{\boldmath$\Phi$}}

\newcommand{\diag}{{\hbox{diag}}}

\newcommand{\trace}{{\hbox{tr}}}
\newcommand{\rank}{{\hbox{rank}}}

\DeclareFontFamily{U}{cmfi}{}
\DeclareFontShape{U}{cmfi}{m}{n}{ <-> cmfi10 }{}
\DeclareSymbolFont{CMFI}{U}{cmfi}{m}{n}

\onecolumn

\begin{document}
  \footernote{To appear in Proc. of IEEE International Symposium on Information Theory (ISIT2010).}

\title{On the Secrecy Degress of Freedom of the\\[-2mm]  Multi-Antenna Block Fading Wiretap Channels}

\author{\authorblockN{Mari Kobayashi, Pablo Piantanida, Sheng Yang}    \\   [1mm]
\authorblockA{Department of Telecommunications, SUPELEC \\
Plateau de Moulon, 91192 Gif-sur-Yvette, France\\
Email: \{mari.kobayashi,pablo.piantanida,sheng.yang\}@supelec.fr}\\[4mm]
\and
\authorblockN{Shlomo Shamai (Shitz)}     \\   [1mm]
\authorblockA{ Technion-Israel Institute of Technology\\
Technion city, Haifa 32000, Israel\\
Email: sshlomo@ee.technion.ac.il}
}

\maketitle

\begin{abstract}
We consider the multi-antenna wiretap channel in which the transmitter wishes to send a confidential message to
its receiver while keeping it secret to the eavesdropper. It has been known that the secrecy capacity of such a channel does not
increase with signal-to-noise ratio when the transmitter has no channel state information (CSI) under mild conditions. Motivated by Jafar's robust interference alignment technique, we study the so-called
staggered multi-antenna block-fading wiretap channel where the legitimate receiver and the eavesdropper have different temporal correlation structures. Assuming no CSI at transmitter,
we characterize lower and upper bounds on the secrecy degrees of freedom (s.d.o.f.) of the channel at hand. Our results show that a positive s.d.o.f. can be ensured whenever two receivers experience different fading variation. Remarkably, very simple linear precoding schemes provide the optimal s.d.o.f. in some cases of interest.
\end{abstract}

\section{Introduction}
In most practical scenarios, perfect channel state information at
transmitter (CSIT) may not be available due to time-varying nature
of wireless channels (in particular for fast fading channels) and
limited resources for channel estimation. However, many wireless
applications must guarantee secure and reliable communication in
the presence of channel uncertainty. In this paper, we consider such a scenario in
the multi-antenna wiretap channel, in which a transmitter wishes to send a confidential message to its receiver
while keeping it secret to the eavesdropper. Although the secrecy capacity of the multi-input multi-output (MIMO) wiretap channel
has been completely characterized for the case of perfect CSIT (see \cite{immse09} and references therein),
the channel at hand is not yet fully understood for the case of imperfect CSIT. Since the complete characterization of the secrecy capacity is generally unknown, most of contributions have focused on secrecy degree of freedom (s.d.o.f) capturing the behavior in high signal-to-noise (SNR) regime.
Modeling the uncertainty at transmitter as a finite set of states, i.e. $\Hc=\{\hv^1,\dots,\hv^{J_1}\}$ and $\Gc=\{\gv^1,\dots,\gv^{J_2}\}$,
the Gaussian MIMO compound wiretap channel has been extensively studied in the literature (see the tutorial \cite{liang2009information} for a complete list of references). In \cite{liangallerton07,liangpimrc08} as well as \cite{kobayashi2009compound}, it has been shown that an achievable s.d.o.f. of the Gaussian compound MIMO wiretap channel collapses as $J_1,J_2$ goes to infinity.
In a recent contribution \cite{khisti2010compound}, Khisti established new upper and lower bounds of the channel at hand and proved that a positive s.d.o.f. which does not depend on $J_1$ and $J_2$ can be achieved under some conditions.
The achievable scheme is based on the so-called real interference alignment, recently proposed in \cite{gou2009degrees,maddah2009degrees}. Although appealing, the upper bound builds on the assumption that any
M vectors taken from the union $\Hc\cup \Gc$ are linearly independent.
A much more robust
interference alignment, applicable to a very general class of sets $\Hc$ and $\Gc$, has been proposed in
\cite{jafar2009exploiting}. In this contribution, Jafar considered the so-called staggered block fading
channel where users have different temporal correlation structures and these structures are known to the transmitter. Remarkably, a simple linear strategy
enables to provide each user a positive d.o.f. with no CSIT in different interference networks.

Inspired by the latter approach, we wish to characterize the achievable s.d.o.f. of the Gaussian multi-input single-output (MISO) wiretap channel by exploiting the opportunity due to the difference in the temporally correlated block fading channels seen by different receivers.
To this end, we consider the so-called staggered MISO block-fading wiretap channel where the legitimate receiver and the eavesdropper have a coherence interval of $T_r$ and $T_e$ blocks and their channels vary with a relative offset of $\Delta$. 
We assume that the transmitter only knows these parameters and the variation of the coherence intervals, whereas each receiver has perfect channel knowledge.
Under a very mild linear independency condition over different temporal vectors,  
we establish lower and upper bounds of the s.d.o.f. for some models of interest.
Our results show that whenever there is a relative difference in the correlation structure between two receivers (either $T_r\neq T_e$ or an offset $\Delta>0$), a positive s.d.o.f can be ensured with no CSIT.
Surprisingly, very simple linear precoding schemes yield the optimal s.d.o.f. in some cases of interest. Similar to the conclusions in \cite{jafar2009exploiting}, the proposed scheme
demonstrates enhanced robustness in terms of available channel state information
accuracy, in comparison to different alignment techniques
proposed for secrecy communications, as in \cite{khisti2010compound,bagherikaram2010secure,koyluoglu2008sdf}.



The rest of the paper is organized as follows. Section \ref{sec:SystemModel} provides the system model and definitions followed by the main result on the s.d.o.f. of the MISO staggered block-fading wiretap channel.
Section \ref{sec:UpperBound}, \ref{sec:LowerBound} provides the sketch of proof for the upper bound, achievable schemes, respectively. Finally Section \ref{sec:conclusions} concludes the paper.

In this paper, we adopt the following notations. We let
$[x]_+=\max\{0,x\}$ and define an indicator function $\onev\{\cdot \}$.
We use
$||\av||, |\Am|, \Am^T, \Am^H, \trace(\Am)$ to denote the norm of a vector $\av$,
the determinant, the transpose, the
hermitian transpose, and the trace of a matrix $\Am$,
respectively. We let $\diag(\av_1,\dots,\av_T)$ denote a block-diagonal matrix whose $t$-th row contains $\av_t$ for all $t$.
We let $\Id_n, \onev_n$ denote a $n\times n$ identity matrix, and a $n\times n$ matrix with unit entries.


\section{System Model and Main Results}\label{sec:SystemModel}
We consider the Gaussian MISO block-fading wiretap channel where the transmitter with $M$ antennas sends a confidential message to the legitimate receiver in the presence of an eavesdropper. The corresponding channel models are given by
\begin{align}\nonumber
    y[\tau] &=  \hv[\tau] \xv[\tau] + n_r[\tau] \\  \label{OriginalModel}
    z[\tau] &=  \gv[\tau] \xv[\tau] + n_e[\tau]
\end{align}
for $\tau=1,\dots,n$,
where $y[\tau], z[\tau]$ denotes the observation at the legitimate receiver, the eavesdropper at channel use $\tau$ associated to $M$-input single-output channel vector $\hv[\tau],\gv[\tau]\in\CC^{1\times M}$, respectively; $n_r[\tau]$ and $n_e[\tau]$ are assumed to be independent and identically distributed (i.i.d.)
complex additive white Gaussian noise~(AWGN) $\sim\Nc_{\Cc}(0,1)$; the input vector $\xv[\tau]\in\CC^{M\times 1}$ is subjected to the average power constraint
$\frac{1}{n}\sum_{\tau=1}^n \trace ( \xv[\tau]  \xv[\tau]^H  )    \leq P.$
We assume that for each $\tau$, $\hv[\tau]$ and $\gv[\tau]$ belong to an arbitrary set of $M$-dimensional vector defined by $\mathscr{H}=\{\av|\;   \gamma_{\min} < ||\av|| < \gamma_{\max} \}$ with some positive and finite $\gamma_{\min}, \gamma_{\max}$.
We further consider the staggered block-fading process $\Cc_{\rm stag}(T_r,T_e,\Delta)$ where $\hv[\tau]$, $\gv[\tau]$ remains constant for a coherence interval of $T_r$, $T_e$ blocks, respectively (i.e. $NT_r$, $NT_e$ channel uses), and change independently with a relative offset of $\Delta$ blocks in arbitrarily manner.



\begin{definition}(code and s.d.o.f. )
 A code $(M_n(P), n,\epsilon_n)$ for the staggered block-fading wiretap channel consists of a message set  $\mc{W}_n=\big\{ 1,\dots, M_n \big\}$ with $w\in \Wc_n$ being uniformly distributed over $\Wc_n$, an stochastic encoder $\big\{ F_S:\mc{W}_n  \longmapsto \mathscr{X}^n  \big\}$ for a random variable $S\in\Sc$ unknown at the decoder, and a legitimate decoder $\big\{ \phi:\mathscr{Y}^n\times \mathscr{H}^n \longmapsto \mc{W}_n  \big\}$  where $\mathscr{H}^n$ is the set of possible channel vectors.
The rate of such code is $\frac{1}{n}\log M_n(P)$ 
and its maximum error probability is defined as
$$
P_e^{(n)}(P)=\max_{w\in \mc{W}_n}\sup\limits_{\hv^n\in \mathscr{H}^n} \Pr\{w \neq \phi(Y^n,\hv^n) | X^n=F_S(w)\}.
$$
Given $0< \epsilon_n,\delta < 1$, a non-negative number $d_s$ is an $\epsilon_n$-achievable s.d.o.f. if for every sufficiently large $n$ there exist codes $(M_n(P), n,\epsilon_n)$  satisfying
 $$
 \liminf\limits_{P \rightarrow \infty } \frac{n^{-1}  \log M_n(P)}{\log P} \geq d_s-\delta
 $$
as well as
\begin{align}\label{ReliabilityConstraint}
& P_e^{(n)}(P)  \leq \epsilon_n \\ \label{PerfectSecrecyConstraint}
&  I(W;Z^n_{\gv^n}) \leq n \epsilon_n, \forall \gv^n\in \mathscr{G}^n
\end{align}
for any $P$ where $\epsilon_n\rightarrow 0$ as $n\rightarrow \infty$. Then, $d_s$ is an achievable s.d.o.f. if it is $\epsilon_n$-achievable for every $0<\epsilon_n<1$.
Then, the supremum of achievable s.d.o.f. is called the optimal s.d.o.f. of the staggered MISO block-fading wiretap channel.
\end{definition}

\begin{prop}\label{proposition}
The largest s.d.o.f. of the MISO staggered block-fading wiretap channel $d_s(T_r,T_e,\Delta)$ is equal to the optimal s.d.o.f. of the Gaussian compound MIMO block wiretap channel defined by
\begin{align}\nonumber
     \yv_{\Hm}[i] &= \Hm \xv[i] + \nv_{r} [i], \forall \Hm\in \Hc\\ \label{EquivalentModel}
     \zv_{\Gm}[i] &= \Gm \xv[i] + \nv_{e}[i], \forall \Gm\in \Gc
\end{align}
for $i=1,\dots,N$ where $\yv_{\Hm}[i],\zv_{\Gm}[i]\in\CC^{T\times 1}$ are obtained by stacking (\ref{OriginalModel}) over $T$ blocks such that $n=TN$, $\xv[i]$ denotes the transmit vector of size $MT$ satisfying $\frac{1}{N}\sum_{i=1}^N \trace(\xv[i]\xv[i]^H) \leq TP$.
We particularize the set $(\Hc,\Gc)$ for the two models:
\begin{itemize}
 \item Different coherence interval model ($T_r\neq T_e,\Delta=0$)\footnote{We assume that the least common multiple of the two integers $(T_r,T_e)$ is given by $T_rT_e$.}
  \begin{align} \label{ArbitraryT1T2Model}
   \Hc &=\big\{ \Hm=\diag(\underbrace{\hv_1,...,\hv_1}_{T_r}, ..,\underbrace{\hv_{T_e},..,\hv_{T_e}}_{T_r})|  \,\forall \hv_i\in \mathscr{H}  \big\}\nonumber  \\
   \Gc& =\big\{\Gm=\diag(\underbrace{\gv_1,.., \gv_1}_{T_e}, ..,\underbrace{\gv_{T_r},..,\hv_{T_r}}_{T_e})|  \,\forall \gv_i\in \mathscr{H} \big\}
\end{align}
where we let $T=T_rT_e$.
\item  Offset model ($T_r=T_e=T$ and $0<\Delta<T$)
\begin{align} \label{OffsetModel}
   \Hc &=\big\{ \Hm=\diag(\underbrace{\hv,\dots, \hv}_{T})|  \,\,\,\,\, \forall \hv \in \mathscr{H}  \big\}\nonumber  \\
   \Gc& =\big\{\Gm=\diag(\underbrace{\gv_1,\dots, \gv_1}_{\Delta}, \underbrace{\gv_2,\dots, \gv_2}_{T-\Delta})|  \,\,\,\,\, \forall \gv_i\in \mathscr{H} \big\}.
\end{align}
 \end{itemize}
\end{prop}
This proposition directly follows from the definition of the staggered block-fading wiretap channel and the assumption that the encoder is unaware of the specific channel realizations. We have the following useful property.
\begin{property}\label{property:interleaving}
By interleaving the underlying staggered block fading channel $\Cc^{\rm stag}(m T_r, m T_e,\cdot)$ in such a way that $\Cc^{\rm stag}(T_r, T_e,\cdot)$ are repeated $m$ times, we have
\begin{align}
d_s(m T_r, m T_e,\cdot)= & d_s(T_r,  T_e,\cdot)
\end{align}
for any integer $m>1$. Similarly we have
\begin{align}
d_s(m T, m T,m\Delta)=& d_s( T,  T,\Delta).
\end{align}
\end{property}


From proposition \ref{proposition} and previous results in \cite{csiszar1978bcc} the following theorem can be easily shown.
\begin{theorem} (Coding theorem)
An achievable rate expression for the staggered MISO block-fading wiretap channel is given by
\begin{equation}\label{CompoundSecrecyRate}
R(P) = \sup\limits_{P_{V\Xm}}\, \inf\limits_{(\mb{H},\mb{G})\in \Hc\times\Gc} \frac{1}{T} \big[I(V;Y_\mb{H})- I(V;Z_\mb{G})\big],
\end{equation}
where the supremum is taken over the set of joint probability densities $P_{V\Xm}$ and an auxiliary random variable $V$ satisfies the Markov chain $V \minuso X \minuso (Y_\mb{H}, Z_\mb{G})$ for all $(\mb{H},\mb{G})\in \Hc\times\Gc$. 
Hence the following s.d.o.f.  is achievable
\begin{equation}\label{Def-sdof}
d_s(T_r,T_e,\Delta) = \lim_{P\rightarrow \infty}\frac{R (P)}{\log(P)}.
\end{equation} \vspace{1mm}
\end{theorem}

We use the following assumption to derive upper and lower bounds on the s.d.o.f..
\begin{assumption} \label{assumption:LinearIndependency}
For each coherence interval $T_{\max}=\max\{T_r,T_e\}$ of one receiver,
there exist at least $r=\min(M,L)$ linearly independent vectors of length $M$ where $L$ denotes the number of channel realizations seen by the other receiver. For the different coherent model with $T_r=qT_e$, there are $\min(M,q)$ linearly independent vectors over $\gv_1,\dots,\gv_q$, while for the offset model, $\gv_1$ and $\gv_2$ are linearly independent.
\end{assumption}

The following remarks are in order:
1) Assumption \ref{assumption:LinearIndependency} imposes some linear independency across the temporal realizations either on the set $\Hc$ or $\Gc$. In other words, we may have $\Hc \subseteq \Gc$ or $\Gc \subseteq \Hc$ and this observation will be extensively used in the converse proof; 2) Assumption \ref{assumption:LinearIndependency} does not imply that \emph{any }$r$ vectors taken from $\gv_1,\dots,\gv_L$ are linearly independent. The latter (more constrained) case yields more optimistic results as discussed in Remarks \ref{remark:UB-LinearIndependency} and \ref{remark:LB-LinearIndependency}.


We overview the main result on the s.d.o.f. of the MISO staggered block-fading wiretap channel. Due to a space limitation, we focus on the special case where $T_{\max}=q T_{\min}$ for the difference coherence interval model. The general result is provided in \cite{kobayashi2010stagger}. It turns out that
for the models at hand upper and lower bounds coincide and yield the optimal s.d.o.f. summarized in the following theorem.

\begin{theorem}\label{theorem:achievableSDoF}
The achievable s.d.o.f. of the staggered MISO block-fading wiretap channel $\Cc_{\rm stag}(T_r,T_e,\Delta)$ is given by
\begin{itemize}
\item Different coherent interval model with $T_{\max}=q T_{\min}$
 \begin{align}\label{Optimal-model1}
    d_s(T_r,T_e,\cdot)= \frac{\min\{M,q\}-1}{q}.
\end{align}
\item Offset model with $T_r=T_e=T$ and $0<\Delta<T$
 \begin{align}\label{Optimal-model3}
    d_s(T,T,\Delta)=\frac{\min\{\Delta,T-\Delta \}}{T}.
\end{align}
 \end{itemize}
\end{theorem}
\vspace{1em}

We have the following interpretations on the result.
\begin{remark}
Theorem \ref{theorem:achievableSDoF} reveals that the difference in the block-fading structure between two receivers provides the opportunity to ensure the secrecy. For the different coherent interval model, there is no relative merit for the legitimate receiver to have the faster or slower fading channel but both cases offer the same opportunity to hide the confidential message by appropriate designs (as specified in Section \ref{sec:LowerBound}).
\end{remark}
\begin{remark}
Another pessimistic interpretation of Theorem \ref{theorem:achievableSDoF} is that the perfect secrecy cannot be garanteed when
both receivers have the same coherence interval $T_r=T_e$ and vary synchronously $\Delta=0$. This is contrasted to the recent result by Khisti \cite{khisti2010compound} and strongly lies on our relaxed assumption. It should be remarked that Assumption \ref{assumption:LinearIndependency} does not impose any linear independency over the different states in compound set $\Hc\times \Gc$ thus includes even the worst case where $\hv[\tau]=\gv[\tau]\in \mathscr{H}$ for some $\tau$. Such a case is excluded in \cite{khisti2010compound} under the assumption that any $M$ vectors taken from $\Hc\times \Gc$ are linearly independent.
\end{remark}

\section{Sketch of Proof of Upper Bounds}\label{sec:UpperBound}
This section provides the sketch of proof of the upper bounds. The proof is divided into two main steps. The first step consists of upper bounding the s.d.o.f. as a function of $|\Tc_c|$ where $\Tc_c$ denotes a subset of blocks $\subset\{1,\dots,T\}$ in which the observations at two receivers are not identically distributed (Lemma \ref{lemma:iid}). Then the second step is to identify the smallest $|\Tc_c|$ for each model by taking into account the linear independency condition of Assumption \ref{assumption:LinearIndependency}.

We extensively use the notation by considering the code length $n= T N$. To this end, we rewrite (\ref{EquivalentModel}) as
\begin{align}\nonumber
     \Ym^n_{\Hm} &= \Hm \Xm^n + \Nm_{r}, \;\forall \Hm\in \Hc\\ \label{EquivalentModel-N}
     \Zm^n_{\Gm} &= \Gm \Xm^n + \Nm_{e}, \;\forall \Gm\in \Gc
\end{align}
where we let $\Ym_{\Hm}^n=[\yv_{\Hm}[1],\dots,\yv_{\Hm}[N]]$. We denote the other variables similarly by stacking $N$ channel uses in columns. Further, we define the $t$-th block output, corresponding to the $t$-th row of $ \Ym^n_{\Hm},\Zm^n_{\Gm}$, as
\begin{eqnarray*}
Y_{t,\hv_t}^N & =& \hv_t \Xm_t^N + \nv_{r,t}^N \\ 
Z_{t,\gv_t}^N & =&  \gv_t \Xm_t^N + \nv_{e,t}^N, \;\; t=1,\dots,T
\end{eqnarray*}
where $Y_{\hv_t,t}^N$ denotes a row vector output of $N$ channel uses, $\Xm_t^N=[\xv_t[1],\dots,\xv_t[N]]$ denotes a $M\times N$ matrix with $\xv[i]^H=[\xv_1[i]^H,\dots,\xv_T[i]^H]$.


We wish to show that there exists a code $(M_n(P), n,\epsilon_n)$ for the staggered MISO block-fading wiretap channel such that
\begin{align*} \label{Fano}
& \frac{1}{n} H(W|\Ym_{\Hm}^n) \leq \epsilon_n, \;\Hm\in \Hc, \;\;
 \frac{1}{n} I(W;\Zm^n_\mb{G}) \leq \epsilon_n, \;\forall \Gm\in \Gc
\end{align*}
where $\epsilon_n\rightarrow 0$ as $n\rightarrow \infty$ for each $P$.
Then, the s.d.o.f. $d_s$ necessarily satisfies the conditions \eqref{Optimal-model1}-\eqref{Optimal-model3}. Notice that the Fano inequality is a consequence of the vanishing error probability condition (\ref{ReliabilityConstraint}).

First, we provide a useful lemma which will be applied subsequently to the two models.
\begin{lemma}\label{lemma:iid}
Let us define
$\Tc \subset\{1,\dots,T\}$ as a collection of $t$ such that
$Y_{t,\hv_t}^N$ and $Z_{t,\gv_t}^N$ are identically distributed vectors and its complementary subset $\Tc_c$ satisfying $\Tc \cup \Tc_c =\{1,\dots,T\}$. For any subset $\Tc_c$, we have
\begin{align}\nonumber
   \frac{1}{n} [I(W;Y_{\Hm}^n) - I(W;Z_{\Gm}^n)] \leq \frac{ |\Tc_c|}{T}\max(\log P, 0) + \frac{C}{T}
\end{align}
where $C$ does not depend on $P,N ,\Hm$ and $\Gm$.
\end{lemma}
\begin{proof}
See \cite{kobayashi2010stagger}.
\end{proof}

From Fano's inequality and the secrecy constraint we have
 \begin{align*}  
\frac{1}{n}\log M_n(P) 
& \leq \frac{1}{n}\inf_{(\Hm,\Gm)\in \Hc\times \Gc} [I(W;Y^n_{\Hm})- I(W;Z^n_{\Gm})]+ 2\epsilon_n \\
&\leq \frac{ |\Tc_c|}{T}\max(\log P, 0) + \frac{C}{T} +2\epsilon_n
\end{align*}
where the last inequality follows from Lemma \ref{lemma:iid}. By dividing both side by $\log P$ and letting $P,N\rightarrow\infty$, we have
\begin{align}
d_s  \leq   \liminf_{P,N\rightarrow\infty} \frac{\log M_n(P)}{n\log P}\leq\frac{ |\Tc_c|}{T}
\end{align}
where we used $\epsilon_n\rightarrow 0$ as $N\rightarrow \infty$. Since the s.d.o.f. upper bound is characterized by a single parameter $|\Tc_c|$, it remains to minimize $|\Tc_c|$ for two models. 

\subsection{Different coherent interval model with $T_{\max}=qT_{\min}$}
We assume $T_r=qT_e$. From Property \ref{property:interleaving}, it is sufficient to let $T=T_r=q$ and $T_e=1$, i.e.
\begin{align}\label{q1Model}
    \Hm = \diag(\hv,\dots,\hv), \;\; \Gm=\diag(\gv_1,\dots,\gv_q).
\end{align}
Since $\hv, \gv_1,\dots,\gv_q \in \mathscr{H}$, the feasible set $\Hc\times \Gc$ includes
\begin{align*}
    \Hm = \diag(\gv_t,\dots,\gv_t), \;\; \Gm=\diag(\gv_1,\dots,\gv_q).
\end{align*}
for any $t=1,\dots,q$.
Further, from Assumption \ref{assumption:LinearIndependency}, we can assume without loss of generality that the first $r=\min(M,q)$ vectors $\{\gv_1,\dots,\gv_r\}$ form the $r$-dimensional space such that $\gv_{t}$ for $t= r+1,\dots,q$ is obtained by a linear combination of these $r$ vectors. Hence, $\Hc\times \Gc$ includes
\begin{align}\label{worstcase}
\Hm=\diag(\gv_r,\dots,\gv_r),\;\;
    \Gm = \diag(\gv_1,\dots,\gv_{r-1},\underbrace{\gv_r,\dots,\gv_r}_{q-r+1}).
\end{align}
It readily follows that $Y_{t,\hv_r}^N$ and $Z_{t,\hv_r}^N$ are identically distributed for $t=r,\dots,q$ since both receivers observe the common channel vector $\gv_r$. In other words, we have $\Tc_c=\{1\dots,r-1\}$. Plugging $|\Tc_c|=r-1$ into Lemma \ref{lemma:iid}, the desired result (\ref{Optimal-model1}) follows.
Due to the symmetry, the same holds for $T_r=qT_e$.

\begin{remark}\label{remark:UB-LinearIndependency}
We wish to remark the impact of Assumption \ref{assumption:LinearIndependency} on the result for the case $M<q$. If we assume instead that \emph{any} $M$ set of vectors taken from $\gv_1,\dots,\gv_{q}$ are linearly independent, $\gv_t$ for $t=M+1,\dots,q$ cannot be obtained as a linear combination of $\gv_1,\dots,\gv_M$ in (\ref{worstcase}). This can only increase the upper bound. Let us consider a simple example with $q=mM$ for some integer $m>1$, we can assume without loss of generality that $\gv_1,\dots,\gv_M$ are repeated successively $m$ times. This yields $|\Tc_s|=m(M-1)$ instead of $M-1$, thus multiplies the s.d.o.f. by $m$.
\end{remark}
\subsection{Offset model with $T_e=T_{r}=T$ and $\Delta>0$}
Since $\hv, \gv_1,\gv_2$ belong to $\mathscr{H}$, the feasible set $\Hc\times \Gc$ includes
\begin{align} \label{caseDelta}
    \Hm = \diag(\underbrace{\gv_1,\dots,\gv_1}_T), \;\; \Gm=\diag(\underbrace{\gv_1,\dots,\gv_1}_{\Delta}, \underbrace{\gv_2,\dots,\gv_2}_{T-\Delta}).
\end{align}
for $\Delta>T-\Delta$ \footnote{For $\Delta<T-\Delta$, we let $\Hm=\diag(\gv_2,\dots,\gv_2)$}.
Assumption \ref{assumption:LinearIndependency} imposes that $\gv_1$ and $\gv_2$ in (\ref{caseDelta}) are linearly independent such that for any $t=\Delta+1,\dots,T$ the observations $Y_{t,\hv}^N$ and $Z_{t,\gv_2}^N$ cannot be identically distributed. Similarly, for $\Delta<T-\Delta$, the observations $Y_{t,\hv}^N$ and $Z_{t,\gv_1}^N$ cannot be identically distributed for $t=1,\dots, \Delta$. which Combines both cases, we have $|\Tc_c|= \min(\Delta, T-\Delta)$. Plugging this into Lemma \ref{lemma:iid}, the desired result (\ref{Optimal-model3}) follows.

\section{Achievable Schemes}\label{sec:LowerBound}
In this section, we provide linear precoding strategies that achieve the optimal s.d.o.f.  (\ref{Optimal-model1}) and (\ref{Optimal-model3}).

\subsection{Different coherence interval model with $T_{\max}=qT_{\min}$}

First we consider the case $T_r=qT_e$ corresponding to the channel structure in (\ref{q1Model}) due to Property \ref{property:interleaving}. For each channel use $i=1,\dots,N$, we form the transmit vector given by
\begin{align}\label{AN}
 \xv = \Phim \vv +
  \left[\begin{array}{c}
         \Id_M \\
         \vdots \\
         \Id_M
       \end{array}\right] \wv
\end{align}
where $\vv\in \CC^{q \times 1}$ denotes the useful signal vector transmitted
along a unitary precoder $\Phim\in\CC^{M q \times q}$ and
$\wv\in\CC^{M\times 1}$ denotes the artificial noise which is
independent of $\vv$. We let $\vv$ and $\wv$ be Gaussian distributed
with covariance $\Sm_v$ and $\Sm_w$, respectively.
The above precoder can be considered as a generalization of the artificial noise scheme \cite{goel2008guaranteeing} to the case of no CSIT. The observations are given by
\begin{align}
     \yv &= \Hm
             \Phim \vv +
          \left[
               \begin{array}{c}
                 1 \\
                 \vdots \\
                  1 \\
               \end{array}
             \right] \hv \wv + \nv_r \\
     \zv &= \Gm\Phim\vv + \overline{\Gm} \wv + \nv_e. 
\end{align}
where we let $\overline{\Gm}^T=[\gv_1^T,\dots,\gv_q^T]$.
In order to determine an achievable s.d.o.f., we let $V=\vv, X=\xv$ and consider equal power allocation such that $\Sm_v= \tilde{P} \Id_{q}$ and $\Sm_w=\tilde{P} \Id_{M}$ with $\tilde{P}=\frac{P}{M+1}$. This yields
 \begin{align}\label{I(v;y)}
 I(\vv;\yv_{\Hm})&= \log\frac{|\Id_{T}+\tilde{P} \Hm\Phim\Phim^\Hm\Hm^\Hm + \tilde{P}\|\hv\|^2\onev_q |}{|\Id_q + \tilde{P}\|\hv\|^2\onev_q  |}\\ \label{I(v;z)}
  I(\vv;\zv_{\Gm})&=\log\frac{|\Id_{T}+\tilde{P}\Gm\Phim\Phim^\Hm\Gm^\Hm +\tilde{P}\Gm\Gm^\Hm|}{|\Id_q+\tilde{P}\overline{\Gm}\overline{\Gm}^\Hm|}.
\end{align}
Since the pre-log factor of $\log |\Id+ P\Am|$ is determined by $\rank(\Am)$ as $P\rightarrow\infty$, we examine the rank of each term.
It can be easily shown that as $P\rightarrow\infty$, the rank of the enumerator and the denominator of
(\ref{I(v;y)}) converges to $q$ and one, respectively.
On the other hand, the enumerator of (\ref{I(v;z)}) has rank $q$ while by assumption \ref{assumption:LinearIndependency} the corresponding denominator, i.e. $\overline{\Gm}$, has rank $r=\min\{q,M\}$. Plugging these equations into (\ref{CompoundSecrecyRate}), it can be easily shown that an achievable secrecy rate is given by
\begin{align}\nonumber
  R(P) &\geq\inf_{(\Hm,\Gm)\in (\Hc,\Gc)} \frac{1}{q}[I(\vv;\yv_{\Hm})-I(\vv;\zv_{\Gm})]\\ \nonumber
  & \geq   \frac{(q-1) - (q-r)}{q} \log P+\Theta(P) \\ \label{prelog-AN}
 &= \frac{r-1}{q} \log P+\Theta(P)
\end{align}
where $\frac{\Theta(P)}{\log P}\rightarrow 0$ as $P\rightarrow\infty$.

Next we consider the case $T_e=qT_r$, corresponding to the channel matrices given by
\begin{align}\label{1qModel}
    \Hm = \diag(\hv_1,\dots,\hv_q), \;\; \Gm=\diag(\gv,\dots,\gv).
\end{align}
We form the transmit vector $\xv$ of size $M T$ as
\begin{align}\label{alignedprecoder}
\xv = \left[\begin{array}{c}
        \Id_M \\
        \vdots \\
        \Id_M
      \end{array}\right] \vv
\end{align}
where $\vv\in\CC^{M\times 1}$ is Gaussian with zero mean and covariance $\Sm_v$. This yields the observations given by
\begin{align}
    \yv = 
\overline{\Hm}\vv + \nv_r, \quad
            \zv = \left[
              \begin{array}{cc}
                  1\\
                  \vdots\\
                 1 \\
              \end{array}
            \right] (\gv\vv) + \nv_e.
\end{align}
where we let $\overline{\Hm}^T=[\hv_1^T,\dots,\hv_q^T]$. From assumption \ref{assumption:LinearIndependency}, the resulting channel $\overline{\Hm}$ has rank $r=\min\{q, M\}$, while the
channel seen by the eavesdropper has rank one. Again we consider equal power
allocation $\Sm_v= \frac{P}{M}\Id_{M}$. From (\ref{CompoundSecrecyRate}), an achievable secrecy rate is given by
\begin{small}
\begin{align}\nonumber
 R(P)  &\geq
 \inf_{(\Hm,\Gm)\in (\Hc,\Gc)} \frac{1}{q} \LSB \log\left|\Id_{q}+ \frac{P}{M}\overline{\Hm}  \overline{\Hm}^H \right | - \log \left(1+  \frac{P||\gv||^2}{M} \right)\RSB\\ \label{prelog-aligned}
 &  \geq  \frac{r- 1}{q}\log P + \Theta(P)
\end{align}
\end{small}
where we have $\frac{\Theta(P)}{\log P}\rightarrow 0$ as
$P\rightarrow\infty$. Dividing the both sides of (\ref{prelog-AN}) and (\ref{prelog-aligned}) by $\log P$ and letting $P\rightarrow\infty$, we have the desired result. This completes the achievability.

\begin{remark}
The achievable scheme (\ref{alignedprecoder}) aligns the useful signal at the eavesdropper via a repetition code, while the artificial noise scheme in (\ref{AN}) aligns the self-interference at the legitimate receiver. These schemes can be naturally extended to the general case of $T_r\neq T_e$ by dividing each interval of $T_{\max}$ into a number of staggered block-fading channels.
\end{remark}

\begin{remark}\label{remark:LB-LinearIndependency}
Similarly to Remark \ref{remark:UB-LinearIndependency}, we examine how the underlying linear independency condition impacts the achievable result for the case $M<q$. If we consider a more restrictive condition than Assumption \ref{assumption:LinearIndependency} such that any $2 \leq l\leq M$ set of vectors taken from $\gv_1,\dots,\gv_{q}$ are linearly independent, the interval of length $q$ can be interleaved into a number of smaller staggered block-fading channels $\Cc_{\rm stag}(l, 1, \cdot)$ for $l\leq M$. By applying the linear precoder (\ref{AN}) to the resulting channel $\Cc_{\rm stag}(l, 1, \cdot)$, we can achieve the s.d.o.f. of $d_s(l,1,\dot)$. As a result, an achievable s.d.o.f. requires the optimization over the block size. We have
\begin{equation}
    d_s(q,1,\cdot) = \max_{2 \leq l\leq q}\left(1-\frac{[q]_l}{q}\right) \frac{\min(l,M)-1}{l}
\end{equation}
where the term $\left(1-\frac{[q]_l}{q}\right) $ accounts for the loss due to the remaining blocks $[q]_l$ over which no information is sent.
\end{remark}

\subsection{Offset model}
The main idea is to exploit two different channel realizations experienced by the eavesdropper over an interval of $T$ blocks. We notice
that $T$ can be divided by two parts of length $2t$ and the remaining block of $T-2t$ with $t=\min\{\Delta,T-\Delta\}$. By interleaving, the first part can be cast into $\Cc_{\rm stag}(2t,t,\cdot)$ where we can apply the linear precoding (\ref{AN}). Since no information is sent over the remaining block of $T-2t$, the number of secured streams is given by
\[d_s(2t,t,\cdot)2t= d_s(2,1,\cdot)2t= t\]
which yields the achievable s.d.o.f. of $t/T$.


\section{Conclusions}\label{sec:conclusions}
In this paper, we characterized upper and lower bounds on the s.d.o.f. of the staggered MISO block-fading wiretap channels by focusing on two models of interest.
Our main fining is that even with no knowledge on the channel realizations at transmitter a positive s.d.o.f. can be ensured as long as the legitimate receiver and the eavesdropper experience different temporal block-fading variations. Moreover, it has been shown that simple linear precoding schemes achieve the optimal s.d.o.f. for the cases studied here. Interestingly, these linear schemes can be easily adapted to other secured communication scenarios such as broadcast channels, and hence this remains as a future investigation.

\section*{Acknowledgment}
This work is partially supported by the European Commission in the framework of the FP7 Network of Excellence in Wireless Communications NEWCOM++.

\bibliographystyle{IEEEtran}
\bibliography{secrecy}
\end{document}